# Novel Binary-Addition Tree Algorithm (BAT) for Binary-State Network Reliability Problem

Wei-Chang Yeh
Department of Industrial Engineering and Engineering Management
National Tsing Hua University
P.O. Box 24-60, Hsinchu, Taiwan 300, R.O.C.
yeh@ieee.org

*Abstract* — Network structures and models have been widely adopted, e.g., for Internet of Things, wireless sensor networks, smart grids, transportation networks, communication networks, social networks, and computer grid systems. Network reliability is an effective and popular technique to estimate the probability that the network is still functioning. Networks composed of binary-state (e.g., working or failed) components (arcs and/or nodes) are called binary-state networks. The binary-state network is the fundamental type of network; thus, there is always a need for a more efficient algorithm to calculate the network reliability. Thus, a novel binary-addition tree (BAT) algorithm that employs binary addition for finding all the possible state vectors and the path-based layered-search algorithm for filtering out all the connected vectors is proposed for calculating the binary-state network reliability. According to the time complexity and numerical examples, the efficiency of the proposed BAT is higher than those of traditional algorithms for solving the binary-state network reliability problem.

## 1. INTRODUCTION

The binary-state network is the fundamental structure of various real-life emerging applications, including network transmission problems involving communication [11, 38] or the distribution [13], transportation [5], transformation [20, 21], and/or transmission of power [23, 38], signals [25], liquids, or gases [2]; grid/cloud computing [37]; data mining [35]; Internet of Things [36]; network topology design [17, 34]; and network resilience problems [12]. Hence, in recent years, the binary-state network has been increasingly researched and applied in the planning, design, execution, management, and control for all the aforementioned systems.

The reliability of the binary-state network is the probability that at least one directed path from the source node to the sink node can be found in the network. The network reliability is an effective measure for evaluating the performance and function of networks and has been widely used in recent decades [2, 5, 11-13, 17, 20, 21, 23, 34-38].

Calculating the binary-state network reliability is an NP-hard problem [3, 8, 15]. Many tools and approaches have been introduced to estimate the binary-state network reliability [19, 22], and they can be categorized into the following: 1) direct algorithms [14] and 2) indirect algorithms [4, 9, 22, 27, 33, 39] based on the minimal cuts (MCs) [1, 9, 22, 27] or minimal paths (MPs) [4, 18, 33, 39]. The reason why MPs and MCs are called “minimal” is that an MP/MC is no longer an MP/MC after any of its arcs are discarded [27, 33, 39].

The direct algorithm comprises two main algorithms: the state-space algorithm and the binary-decision diagram (BDD) [6, 14]. As indicated by its name, the state-space algorithm constructs a space containing all the possible state vectors; thus, it is less efficient than the other algorithms [14]. The BDD requires extensive coding skill using complex data structures, and there is no public source code for users and researchers to download and compare [10]. Hence, indirect algorithms based on MCs or MPs are more popular than the direct algorithms [4, 9, 18, 22, 27, 33, 39].

For the indirect algorithms [1, 4, 9, 22, 27, 33, 39], there are two major steps in calculating the binary-state network reliability based on MPs/MCs: 1) find all the MPs/MCs [1, 4, 9, 22, 27, 33, 39]; 2) calculate the binary-state network reliability in terms of the MPs/MCs using the inclusion–exclusion technique (IET) [10, 30] or the sum-of-disjoint product method (SDP) [6, 28, 32, 39]. The two aforementioned stages both involve NP-hard problems [13–15].

Searching for entire MPs is still far less efficient than searching for MCs in general binary-state networks [16]. To calculate the network reliability, the most efficient algorithm was proposed by Yeh in terms of the MPs in acyclic binary-state networks. Yeh’s algorithm was a node-based Depth-first search (DFS) [33] and was the first algorithm to find all the MPs with the same time complexity that was observed for previously reported algorithms in finding all the MCs [16].

The size of the modern network is increasing from year to year, and the application of network reliability is accordingly becoming broader and more flexible [7]. Hence, there is always a need to develop a more efficient algorithm to calculate the exact binary-state network reliability [7, 22]. The objective of this study was to develop a direct algorithm for calculating the binary-state network reliability that does not use MPs and MCs and is significantly more efficient than existing MP and MC algorithms [16, 19].

This remainder of this paper is organized as follows. Acronyms, notations, nomenclatures, and assumptions are presented in Section 2. Section 3 reviews the current indirect methods, including the MP and MC algorithms [1, 4, 9, 22, 27, 33, 39], which find all MPs and MCs, respectively, and the IET [10, 30] and SDP [28, 32, 39], which calculate the reliability in terms of the found MPs and MCs. The innovative parts of the proposed binary-addition tree algorithm (BAT) are presented in Section 4, including the binary-addition tree for finding all the state vectors, the path-based layered-search algorithm (PLSA) to select connected state vectors, the reduction method to reduce the number of state vectors, the connectivity verifications to reduce the runtime, and the method for calculating the probability of each connected vector. The total of these probabilities is the reliability. Section 5 presents the pseudocode, a step-by-step example, and the time complexity of the proposed BAT in detail. Additionally, a computational experiment was performed to compare the performance between the proposed BAT and the best-known indirect method [10], as described in Section 5. Concluding remarks on this study are presented in Section 6.

## 2. ACRONYMS, NOTATIONS, NOMENCLATURES, AND ASSUMPTIONS

Relevant acronyms, notations, assumptions, and nomenclatures are presented in this section.

### 2.1 Acronyms

MP/MC: minimal path/cut

BAT: binary-addition tree algorithm

LSA: layered-search algorithm [26]

PLSA: path-based layered-search algorithm

DFS: depth-first-search method

IET: inclusion–exclusion technology

SDP: sum-of-disjoint products method

**2.2 Notations**

|•|: number of elements in set •

$n$, $m$: numbers of nodes and arcs, respectively

$V$: node set $V = \{1, 2, \ldots, n\}$

$E$: arc set $E = \{a_1, a_2, \ldots, a_m\}$

$e_{i,j}$: directed arc between nodes $i$ and $j$

$a_k$: $k^{\text{th}}$ directed arc in $E$

**D**: state distributions of arcs list the success probability of each arc

$P(\bullet)$: success probability of event •

$G(V, E)$: A graph with $V$, $E$, the source node 1, and the sink node $n$. For example, Fig. 1 shows a graph with $V = \{1, 2, 3, 4\}$ and $E = \{a_1 = e_{1,2}, a_2 = e_{1,3}, a_3 = e_{2,3}, a_4 = e_{2,4}, a_5 = e_{3,4}, a_6 = e_{2,1}, a_7 = e_{3,1}, a_8 = e_{4,2}, a_9 = e_{4,3}\}$, the source node 1, and the sink node 4.

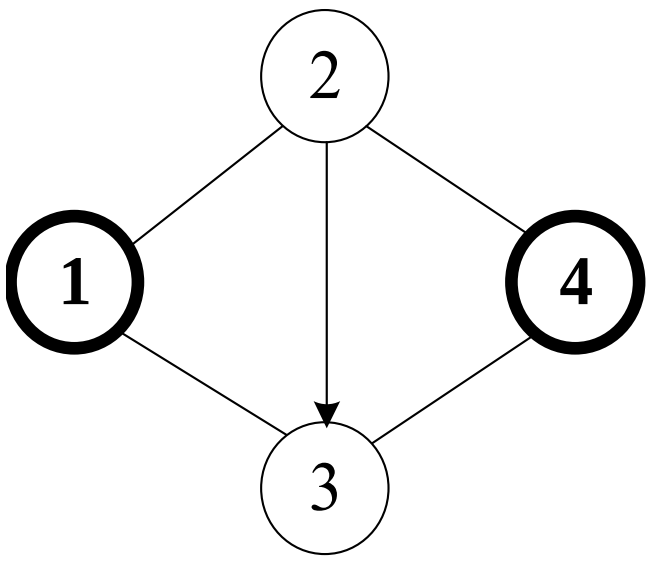

**Figure 1.** Example network

$G(V, E, \mathbf{D})$: A binary-state network with graph $G(V, E)$ and state distributions **D**. For example, Fig. 1 becomes a binary-state network $G(V, E, \mathbf{D})$ after the addition of **D**, which is presented in Table 1.

**Table 1.** Arc state distributions in Fig. 1

| $e$ | Pr($e$) | $e$ | Pr($e$) |
|---|---|---|---|
| $a_1 = e_{1,2}$ | 0.8 | $a_6 = e_{2,1}$ | 0.9 |
| $a_2 = e_{1,3}$ | 0.9 | $a_7 = e_{3,1}$ | 0.8 |
| $a_3 = e_{2,3}$ | 0.7 | | |
| $a_4 = e_{2,4}$ | 0.8 | $a_8 = e_{4,2}$ | 0.7 |
| $a_5 = e_{3,4}$ | 0.9 | $a_9 = e_{4,3}$ | 0.8 |

$R$: reliability of binary-sate network $G(V, E, D)$

$X$: state vector whose $i^{\text{th}}$ coordinate represents the state of $a_k$ for $k = 1, 2, \ldots, m$

$X(a_i)$: state (value) of the $a_i$ (the $i^{\text{th}}$ coordinate) in $X$

$\Pr(X(a_i))$: occurrence probability of $a_i$ when its state is $X(a_i)$

$\Pr(X)$: $\Pr(X) = \Pr(X(a_1))\cdot\Pr(X(a_1))\cdot\ldots\cdot\Pr(X(a_m))$

$\Pr^{-}(X)$: $\Pr^{-}(X) = \Pr(\{X \mid \text{for all } X^* \text{ with } X^* \le X\})$

$G(X)$: The subgraph corresponding to state vector $X$ such that $G(X) = G(V, \{a\in E \mid \text{for all } a \text{ with } X(a) = 1\})$; e.g., the graph of $G(X)$ is depicted in Fig. 2, where $X = (1, 1, 1, 1, 1, 0, 0, 0, 0)$.

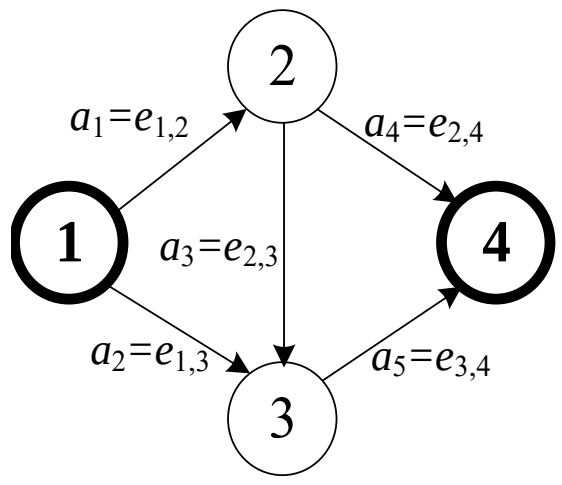

**Figure 2.** $G(X)$ and $X = (1, 1, 1, 1, 1, 0, 0, 0, 0)$ in Fig. 1.

$n_p$: The number of arcs in the shortest paths from nodes 1 to $n$; e.g., $n_p = 2$ because $\{a_1, a_4\}$ and $\{a_2, a_5\}$ are shortest paths.

$n_c$: The number of arcs in minimum cuts between node 1 and node $n$; e.g., $n_c = 2$ because $\{a_1, a_2\}$ and $\{a_4, a_5\}$ are minimum cuts.

$A \le B$: $A(a_i) \le B(a_i)$ for all $i = 1, 2, \ldots, m$, e.g., $(1, 2, 3, 4, 5) \le (1, 2, 3, 4, 6)$

$A < B$: $A \le B$ for all $i = 1, 2, \ldots, m$ and $A(a_j) < B(a_j)$ for at least one $j = 1, 2, \ldots, m$, e.g., $(1, 2, 3, 4, 5) < (1, 2, 3, 4, 6)$

Max, Min: Maximal element and the minimal element, respectively, e.g., Max({3, 2}) = 3 and Min({1, 5}) = 5

## 2.3 Nomenclatures

Reliability: The success probability that there is one direct path from node 1 to node $n$

MP: A path from node 1 (the source node) to node $n$ (the sink node) is an MP if any of its proper subsets is not an MP; e.g., $\{a_1, a_3, a_5\}$ is an MP from node 1 to node $n$ in Fig. 1.

MC: A cut between node 1 (the source node) and node $n$ (the sink node) is an MC if any of its proper subsets is not an MC; e.g., $\{a_1, a_2\}$ is an MC in Fig. 1.

Connected vector: A state vector $X$ is connected if nodes 1 and $n$ are connected and there is at least one MP in $G(X)$.

### 2.4 Assumptions

1. All nodes of the network are completely reliable.
2. There are only two situations in each arc of the binary-state network.
3. The network is connected and has no parallel arcs or loops.
4. The success probability of each arc is statistically independent according to a given distribution.

## 3. REVIEW OF INDIRECT ALGORITHMS

MP and MC algorithms are the main types of indirect algorithms for calculating the network reliability [1, 4, 9, 22, 27, 33, 39]. In the MP algorithms [4, 33, 39] and MC algorithms [1, 9, 22, 27], we search for the complete MP set and the complete MC set first, respectively. Then, the IET [10, 30] or SDP [28, 32, 39] is used to calculate the final reliability in terms of the MP set obtained using the MP algorithms [4, 33, 39] or the MC set obtained using the MC algorithms [1, 9, 22, 27]. The performance of the proposed BAT is validated via comparison with these algorithms. Hence, MPs, MCs, the IET [10, 30], and SDP [28, 32, 39] are introduced briefly in this section.

### 3.1 MPs

An MP is a directed simple path from node 1 to node $n$, and an MP is no longer a path if any arc is removed. There are different methods to search for MPs, including the depth-first search (DFS) [33], the heuristic method [29], and the universal generating function method [15, 16, 31].

The number of MPs is $O(2^m) = O(2^{n^2})$ [33, 39]. The time complexity to search for all MPs was Min$\{\binom{m}{n}O(2^n), O(2^m)\}$ until Yeh proposed a node-based DFS for acyclic networks [33]. Yeh's node-based DFS introduced the novel concept that an MP can be represented by an ordered node subset, reducing its time complexity to $O(2^n)$ [33]. Note that $O(m) = O(n^2)$.

Yeh's node-based DFS is the most efficient one among all the MP-based algorithms [33, 39]. However, similar to all MP-based algorithms, Yeh's node-based DFS [33] finds all MPs only and invokes to implement the IET [10, 30] or SDP [28, 32, 39] to calculate the final reliability.

### 3.2 MCs

A cut is an arc subset whose removal from the network causes nodes 1 and $n$ to be disconnected. Analogous to the definition of an MP, any proper subset of an MC is not an MC. Each MC can be treated as an unordered node subset, such that the number of MCs is $O(2^n)$ [1, 9, 22, 27], but each MP can only be an ordered node subset [33, 39]. Hence, the number of MCs is theoretically smaller than the number of MPs.

Among all the MC algorithms, the MCV-DFS proposed by Yeh outperforms others in searching for all MCs in general networks [1, 9, 22, 27]. Yeh's MCV-DFS is based on the MCV concept, which is the related node subset including node 1 after the removal of an MC [27]. The time complexity to search for all MCs is $O(2^n)$ in MCV-DFS [27].

### 3.3 Inclusion–Exclusion and Sum-of-Disjoint Product Methods

In the indirect algorithms [1, 4, 9, 22, 27, 33, 39], the last step is to use all the found MPs or MCs to calculate the network reliability [27, 33, 39]. Let $p_1, p_2, \ldots, p_\pi$ be all the MPs. The IET [10, 30] and SDP [6, 28, 32, 39] are given by Eqs. (1) and (2), respectively:

$$R=\sum_{i=1}^{\pi}\Pr(p_i)-\sum_{j=2}^{\pi}\sum_{i=1}^{j-1}\Pr(p_i\cap p_j)+\sum_{k=3}^{\pi}\sum_{j=2}^{k-1}\sum_{i=1}^{j-1}\Pr(p_i\cap p_j\cap p_k)+\ldots+(-1)^{\pi}\Pr(p_1\cap\ldots\cap p_\pi), \quad (1)$$

$$R=\Pr(\{X\mid p_1\le X\})+\Pr(\{X\mid p_2\le X \text{ and } X<p_1\})+\ldots+\Pr(\{X\mid p_\pi\le X, X<p_1,\ldots,X<p_{\pi-1}\}), \quad (2)$$

where

$$\Pr(p_i)=\Pr(\{X\mid \text{for all } X \text{ with } p_i\le X\}). \quad (3)$$

Similarly, we have the IET [10, 30] and SDP [28, 32, 39] based on all the found MCs [9, 27]. More detailed information on the these IET and SDP can be found in recently published works [10, 32].

## 4. INNOVATION PARTS IN BINARY-ADDITION TREE ALGORITHM

In the proposed BAT, all the state vectors are found first. Then, the connected vectors are filtered out from these state vectors. The network reliability is obtained by calculating the occurrent and summed up probabilities of the connected state vectors. Hence, a simple binary-addition tree, the PLSA, and reduction methods are used to find all the state vectors, verify the state vectors, and reduce the number of possible state vectors and the number of state vectors to be verified, respectively. Additionally, a method to calculate the probabilities of the connected vectors is presented in this section.

### 4.1 Binary-Addition Tree

In the proposed BAT, a simple binary-addition tree is used to generate all the possible state vectors before the binary-state network reliability is calculated. A state vector is a vector for which the value of any of its coordinates is either 0 or 1 and represents the state of the related arc. In the BAT, each new state vector is generated after adding 1 to the current state vector, e.g., $X$, by treating it as a binary number such that the value of its $i^{\text{th}}$ digit is equal to $X(a_i)$. The orders of the arcs in all the state vectors are identical, such that no two state vectors have the same values.

The overall procedure of the binary-addition tree for the proposed BAT is described by the pseudocode below.

**Algorithm for the binary-addition tree:** Verify whether there is a directed (1, $n$)-path in $G(X)$ for the state vector $X$.

**Input:** $G(V, E)$.

**Output:** All possible state vectors without duplications.

**STEP A0.** Let SUM = 0, $k = 1$, and $X_1 = X$ be a zero vector with $m$ coordinates.

**STEP A1.** Let $i = m$.

**STEP A2.** If $X(a_i) = 0$, let $X(a_i) = 1$, $k = k + 1$, $X_k = X$, SUM = SUM + 1, and go STEP A5.

**STEP A3.** Let $X(a_i) = 0$. If $i > 1$, let $i = i - 1$ and go STEP A2.

**STEP A4.** If SUM = $m$, halt and $X_1$, $X_2$, …, $X_k$ are all possible state vectors. Otherwise, go to STEP A1.

In STEP A0, the first state vector, i.e., $X_1$, in the proposed BAT is a zero vector, and all the other vectors are generated by adding 1 according to the binary addition in sequence, as shown in a loop from STEP A1 to STEP A3. The foregoing procedure only stops when all the coordinates are 1, i.e., SUM = $m$ in STEP A4.

For example, for the graph shown in Fig. 2, $X_1$ is a 5-tuple zero state vector, i.e., $X_1$ = (0, 0, 0, 0, 0). From the binary addition, we have

$$00000 + 1 = 00001, \tag{4}$$

$$00001 + 1 = 00010, \tag{5}$$

$$00001 + 1 = 00011, \tag{6}$$

$$00011 + 1 = 00100. \tag{7}$$

after using the above pseudocode for the first five vectors. Hence, from Eqs. (4)–(7), $X_2$ = (0, 0, 0, 0, 1), $X_3$ = (0, 0, 0, 1, 0), $X_4$ = (0, 0, 0, 1, 1), and $X_5$ = (0, 0, 1, 0, 0). Similarly, we have all the state vectors. The state vectors obtained from the binary-addition tree using the above pseudocode are presented in Table 2.

**Table 2.** All the $X_i$ values obtained in the proposed BAT.

| $i$ | $B_i$ | $X_i$ | Connected? | $i$ | $B_i$ | $X_i$ | Connected? |
|---|---|---|---|---|---|---|---|
| 1 | 00000 | (0, 0, 0, 0, 0) | N | 17 | 10000 | (1, 0, 0, 0, 0) | N |
| 2 | 00001 | (0, 0, 0, 0, 1) | N | 18 | 10001 | (1, 0, 0, 0, 1) | N |
| 3 | 00010 | (0, 0, 0, 1, 0) | N | 19 | 10010 | (1, 0, 0, 1, 0) | Y |
| 4 | 00011 | (0, 0, 0, 1, 1) | N | 20 | 10011 | (1, 0, 0, 1, 1) | Y |
| 5 | 00100 | (0, 0, 1, 0, 0) | N | 21 | 10100 | (1, 0, 1, 0, 0) | N |
| 6 | 00101 | (0, 0, 1, 0, 1) | N | 22 | 10101 | (1, 0, 1, 0, 1) | Y |
| 7 | 00110 | (0, 0, 1, 1, 0) | N | 23 | 10110 | (1, 0, 1, 1, 0) | Y |
| 8 | 00111 | (0, 0, 1, 1, 1) | N | 24 | 10111 | (1, 0, 1, 1, 1) | Y |
| 9 | 01000 | (0, 1, 0, 0, 0) | N | 25 | 11000 | (1, 1, 0, 0, 0) | N |
| 10 | 01001 | (0, 1, 0, 0, 1) | Y | 26 | 11001 | (1, 1, 0, 0, 1) | Y |
| 11 | 01010 | (0, 1, 0, 1, 0) | N | 27 | 11010 | (1, 1, 0, 1, 0) | Y |
| 12 | 01011 | (0, 1, 0, 1, 1) | Y | 28 | 11011 | (1, 1, 0, 1, 1) | Y |
| 13 | 01100 | (0, 1, 1, 0, 0) | N | 29 | 11100 | (1, 1, 1, 0, 0) | N |
| 14 | 01101 | (0, 1, 1, 0, 1) | Y | 30 | 11101 | (1, 1, 1, 0, 1) | Y |
| 15 | 01110 | (0, 1, 1, 1, 0) | N | 31 | 11110 | (1, 1, 1, 1, 0) | Y |
| 16 | 01111 | (0, 1, 1, 1, 1) | Y | 32 | 11111 | (1, 1, 1, 1, 1) | Y |

In Table 2, the values of $a_1$, $a_2$, $a_3$, $a_4$, and $a_5$ are presented in the coordinates 1, 2, …, 5 of each vector, respectively. Columns $B_i$ show the related binary code of $X_i$. The columns entitled "Connected?" indicate whether the related state vector $X$ in $G(X)$ is connected or disconnected; e.g., $X_{16}$ = (0, 1, 1, 1, 1) is connected implies that nodes 1 and $n$ in $G(X_{16})$ are connected. Details regarding how to verify the state vectors are presented in Section 4.2.

In the DFS tree of Fig. 3, let $b_i$ be the $i^{\text{th}}$ branch (counted from the leftmost leaf to the rightmost leaf), where $b_i$(Forghani-Elahabad and Kagan) represents the $j^{\text{th}}$ coordinate in $b_i$ denoted the state of arc $a_j$, $a_1 = e_{1,2}$, $a_2 = e_{1,3}$, $a_3 = e_{2,3}$, $a_4 = e_{2,4}$, $a_5 = e_{3,2}$, and $i$ = 1, 2, …, 32. Because each coordinate is either 0 or 1 and there are five coordinates, we have $2^5$=32 combinations of different $b$ as shown in Fig. 3.

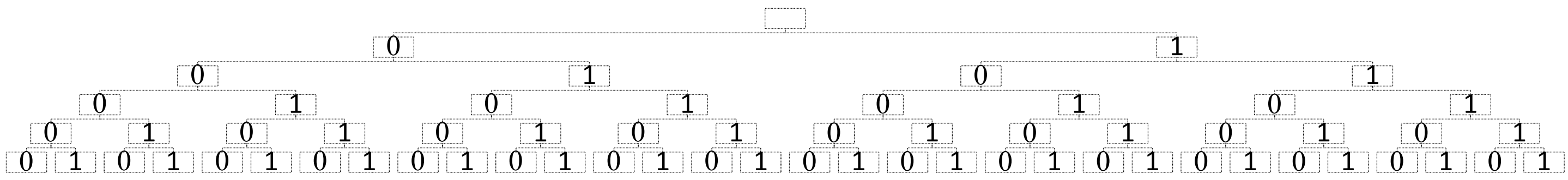

**Figure 3**. DFS tree of Fig. 2.

According to Table 2 and Fig. 3, all the state vectors obtained from the proposed BAT can be obtained without duplications from the DFS tree and vice versa. Thus, all the possible state vectors are found without duplications from the proposed BAT.

## 4.2 PLSA

In the proposed BAT, the connectivity of each state vector must be verified before calculating the network reliability. To achieve this goal efficiently, the PLSA is proposed as follows.

Each state vector represents a sub-network of the original binary-state network in the BAT; e.g., the sub-networks represented by $X_{15}$ = (0, 1, 1, 1, 0) and $X_{16}$ = (0, 1, 1, 1, 1) are depicted in Figs. 4a and 4b, respectively.

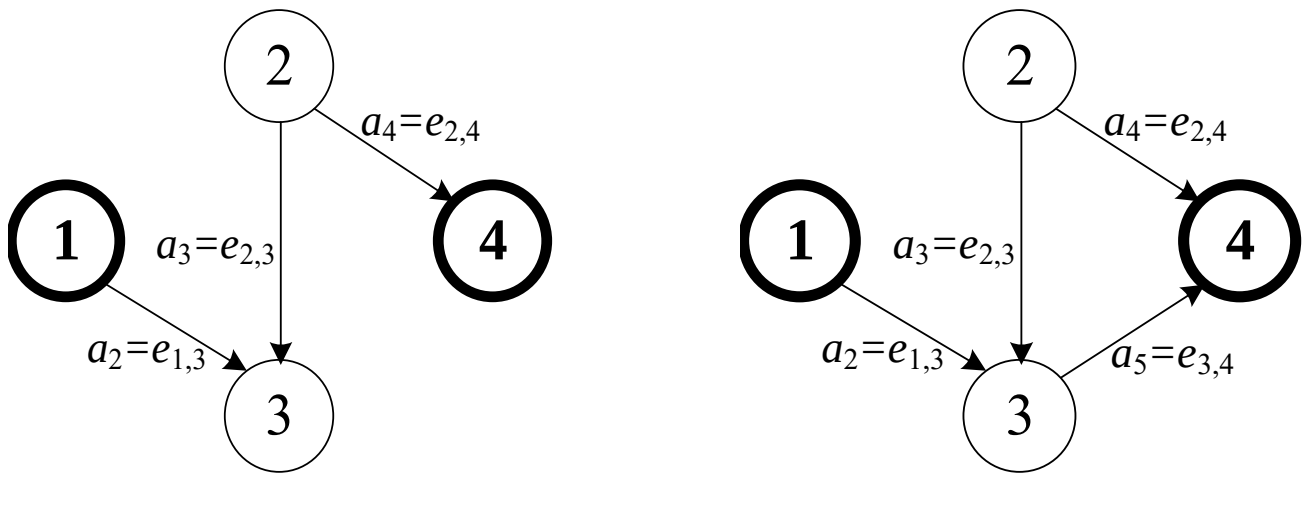

(a) $G(X_{15})$ (b) $G(X_{16})$

**Figure 4.** Sub-networks represented by $X_{15} = (0, 1, 1, 1, 0)$ and $X_{16} = (0, 1, 1, 1, 1)$.

According to the foregoing discussion, we can determine whether nodes 1 and $n$ are connected in the subgraph related to the state vector to test the connectivity of this vector.

The original layered-search algorithm (LSA) was proposed in [26] to search for all $d$-MPs in acyclic networks. Owing to its simplicity and efficiency, the LSA was revised (resulting in the PLSA) to verify whether a direct (1, $n$)-path exists in the related graph of each state vector in the proposed BAT.

In the proposed PLAS, there is an ordered vertex subset $V^*$ that stores these nodes that have been reached from node 1 in sequence. Moreover, $V^* = L_1 \cup L_2 \cup \ldots \cup L_k$, where layer $L_1 = \{1\}$ and layer $L_i = \{ v \in [V - (L_1 \cup L_2 \cup \ldots \cup L_{i-1})] \mid \text{for all } e_{\alpha,v} \in E \text{ and } \alpha \in L_{i-1}\}$ for $i = 2, 3, \ldots, k$. It is trivial that there is at least a directed (1, $n$)-path if $n \in L_i$ and there is at least a (1, $n$)-cut if $L_i = \varnothing$ in $G(V^*)$. The pseudocode of the proposed PLSA is as follows.

**Algorithm PLAS:** Determine whether there is a directed (1, $n$)-path in $G(X)$ for the state vector $X$.

**Input:** A state vector $X$ in the binary-addition tree

**Output:** Whether the state vector $X$ is connected or disconnected

**STEP P0.**Let $i = 2$ and $V^* = L_1 = \{1\}$

**STEP P1.**Let $L_i = \{ v \notin V^* \mid \text{for all } e_{\alpha,v} \in E \text{ and } \alpha \in V^*\}$.

**STEP P2.**If $n \in L_i$, there is a directed (1, $n$)-path in $G(X)$. Stop.

**STEP P3.**If $L_i = \varnothing$, there is a (1, $n$)-cut in $G(X)$. Halt.

**STEP P4.**Let $V^* = V^* \cup L_i$, $i = i + 1$, and go to STEP P1.

For example, in Fig. 2, the procedure to determine whether $X = (1, 1, 1, 1, 1)$ is a connected state vector based on the proposed PLSA is presented in Table 3.

**Table 3.** Process from the proposed PLSA for Fig. 2.

| $i$ | $L_i$ | $L_{i+1}$ | $V^*$ | Remark |
|---|---|---|---|---|
| 1 | {1} | {2, 3} | {1, 2, 3} | |
| 2 | {2, 3} | {4} | {1, 2, 3, 4} | a directed (1, $n$)-path found |

The proposed PLSA can be used only to test the connectivity of a state vector. Because there are at most $n$ nodes to be found in $V^*$, the time complexity of the proposed PLSA is $O(n)$ in verifying whether a state vector is connected.

## 4.3 Reduction Methods

In the proposed BAT, each vector has all the states of directed arcs. Because of the NP-hard characteristic, the runtime is decreased to half of the original value if the number of directed arcs is reduced by 1. Additionally, the performance is improved if the number of connectivity verifications is reduced. To achieve these two goals, two arc reductions are introduced for increasing the efficiency of the proposed BAT.

### **4.3.1** Reduce Number of Arcs

A directed (1, $n$)-path from the sink node to the source node cannot exist. Hence, to reduce the computational burden, all arcs that are impossible to use are removed from the original network, with no effect on the final reliability, e.g., all arcs from node $n$ (the sink node) to node 1 (the source node).

For example, the reliabilities of Figs. 1 and 2 are identical, and Fig. 1 can be reduced to Fig. 2 after removing $e_{2,1}$, $e_{3,1}$, $e_{4,2}$, and $e_{4,3}$ in calculating the reliability. Subsequently, the number of directed arcs is reduced from nine to five, and the runtime can theoretically be $2^4$ times shorter than the original value, as the number of possible state vectors is reduced from $2^9$ to $2^5$.

### **4.3.2** Reduce Number of Connectivity Verifications

Another important method for improving the proposed BAT is to reduce the number of verifications of whether nodes 1 and $n$ are connected in $G(X)$ for each vector $X$.

Let $n_p$ be the directed shortest (1, $n$)-path and $n_c$ be the minimum (1, $n$)-cut in $G$. No directed (1, $n$)-path or (1, $n$)-cut has a number of arcs smaller than $|p|$ or $|c|$, respectively. Hence, nodes 1 and $n$ are disconnected and connected in $G(X)$ if the numbers of state 1s and 0s in the state vector $X$ are smaller than $n_p$ and $n_c$, respectively, for $i = 1, 2, \ldots, m$. The number of zero states is less than $n_c$ equivalent to that the number of state ones is larger than $m-n_c$ in $X$.

For example, $\{a_1, a_2\}$ and $\{a_1, a_4\}$ are one of the minimum (1, $n$)-cuts and one of the shortest directed (1, $n$)-paths in Fig. 2, respectively. We have $n_p = n_c = 2$. $X_1$, $X_2$, $X_3$, $X_5$, $X_9$, and $X_{17}$ are all disconnected, as their numbers of state 1s are smaller than $n_p = 2$. $X_{24}$, $X_{28}$, $X_{30}$, $X_{31}$, and $X_{32}$ are all connected, as their numbers of state 0s are smaller than $n_c = 2$. Because the 11 aforementioned state vectors do not need to be verified, the number of verifications can be reduced from $2^5 = 32$ to 21 as shown in Table 2.

Moreover, after using only the two aforementioned bounds, the number of verifications is reduced at most by

$$\sum_{j=0}^{n_p-1} 2^j \binom{m}{j} + \sum_{j=m-n_c+1}^{m} 2^j \binom{m}{j}. \tag{8}$$

### 4.4 Calculation of Binary-State Network Reliability Using Connected State Vectors

In the proposed BAT, after all the state vectors are obtained, the final step is to calculate the binary-state network reliability according to these state vectors.

The occurrent probability of state vector $X$ in the BAT is the product of the reliabilities of the working arcs and the un-reliabilities of the failed arcs, i.e.,

$$P\,r(X_i) = \prod_{j=1}^{m} Pr(X_i(a_j)). \tag{9}$$

All the possible state vectors are found in the BAT. For all the state vectors (connected or disconnected), we have

$$\sum_{\forall i} P\,r(X_i) = \sum_{\forall i} \prod_{j=1}^{m} Pr(X_i(a_j)) = 1. \tag{10}$$

The reliability $R$ of the binary-state network is the summation of all the occurrent probabilities of these connected state vectors, e.g., $X$, such that nodes 1 and $n$ are connected in $G(X)$:

$$R = \sum_{\forall X} P\,r(X). \tag{11}$$

For example, let $p_i$ and $q_i = 1 - p_i$ be the success and failure probabilities of $a_i$ in Table 2, respectively. We have $Pr(B_1) = q_1q_2q_3q_4q_5$ and $Pr(B_2) = q_1q_2q_3q_4p_5$, $Pr(B_1) + Pr(B_2) + \ldots + Pr(B_{32}) =$

1, and $R = Pr(B_{10}) + Pr(B_{12}) + Pr(B_{14}) + Pr(B_{16}) + Pr(B_{19}) + Pr(B_{20}) + Pr(B_{22}) + Pr(B_{23}) + Pr(B_{24}) + Pr(B_{26}) + Pr(B_{27}) + Pr(B_{28}) + Pr(B_{30}) + Pr(B_{31}) + Pr(B_{32})$.

## 5. PROPOSED BAT

The procedure for the proposed BAT for generating all the possible connected state vectors and evaluating the reliability according to these connected vectors in binary-state networks is presented in Section 5.2. A demonstration is presented in Section 5.1. Additionally, the performance of the proposed BAT was compared with that of the best-known algorithm for 20 benchmark problems, as described in Section 5.3.

### 5.1 Pseudocode and Time Complexity of BAT

The pseudocode of the proposed BAT based on the components proposed in Section 4 is presented as follows.

**Input:** A binary-state network $G(V, E, \mathbf{D})$

**Output:** The reliability $R$

**STEP 0.** Remove the arcs from node $n$, to node 1, and find $n_p$, $n_c$, and $m$ (the number of directed arcs in the new graph).

**STEP 1.** Let $R = k = 0$, $X(a_i) = 0$ for all $i = n_p+1, n_p+2, \ldots, m$, $X(a_i) = 1$ for all $i = 1, 2, \ldots, n_p$, SUM $= n_p$, and go to STEP 6.

**STEP 2.** Let $i = m$.

**STEP 3.** If $X(a_i) = 0$, let $X(a_i) = 1$ and SUM = SUM + 1, and go STEP 5.

**STEP 4.** Let $X(a_i) = 0$. If $i > 1$, let $i = i - 1$, and go STEP 3.

**STEP 5.** If SUM $< n_p$, go to STEP 2.

**STEP 6.** If SUM $> m - n_c$, go to STEP 8.

**STEP 7.** If nodes 1 and $n$ are disconnected in $G(X)$ after the PLSA is used to verify its connectivity, go to STEP 2.

**STEP 8.** Let $k = k + 1$, $X_k = X$, and $R = R + R(X)$.

**STEP 9.** If SUM = $m$, $X_1$, $X_2$, …, $X_k$ are all the possible state vectors, and $R$ is the final reliability; halt. Otherwise, go to STEP 2.

STEP 0 is based on Section 4.3.1 and is for reducing the number of directed arcs. STEPs 1–8 mainly implement the binary-addition tree discussed in Section 4.1 by integrating the PLSA and the reduction methods mentioned in Section 4.3. STEP 0 is based on the arc-number reductions discussed in Section 4.3.1. STEPs 5 and 6 are from Section 4.3.2 and are for reducing the number of connectivity verifications. STEP 7 executes the proposed PLSA to verify the connectivity of these vectors that have not been tested in STEPs 5 and 6. STEP 8 indexes each found connected vector and calculates its reliability.

In the network-reliability method, the time complexity is always useful for comparing the performance among all the related algorithms. The algorithm with the best time complexity is theoretically the most efficient [8, 22].

The time complexity of the proposed BAT is $O((n+m)2^m) = O((n^2 \cdot 2^{n^2})$, where $O(n)$, $O(m) = O(n^2)$, and $O(2^m) = O(2^{n^2})$ are the time complexities to implement the proposed PLSA to verify the connectivity of each state vector from Section 4.2, to calculate the probability of each connected state vector based on Section 4.4, and to calculate the total number of state vectors obtained from Section 4.1 (e.g., there are 32 state vectors in Table 2), respectively.

The numbers of MPs and MCs are $O(|P|) = O(2^m) = O(2^{n^2})$ [33, 39] and $O(|C|) = O(2^n)$ [1, 9, 22, 27], respectively. The time complexities of both the IET and the SDP are $O(m2^{\mathrm{N}})$, where N represents the number of elements or events [30]. Therefore, the MP algorithms and MC algorithms have time complexities of $O(m \cdot 2^{|P|}) = O(n^2 \cdot 2^{2^{n \cdot n}})$ [10, 33, 39] and $O(m2^{|C|}) = O(n^2 \cdot 2^{2^n})$ [9, 10, 27], respectively.

According to the foregoing discussion, because $O(n^2 \cdot 2^{n^2}) << \mathrm{Min}\ \{O(n^2 \cdot 2^{2^{n \cdot n}}), O(n^2 \cdot 2^{2^n})\} = O(n^2 \cdot 2^{2^n})$, the proposed BAT is faster than the current MP algorithms and MC algorithms from the viewpoint of the time complexity. Furthermore, the practical performance of the proposed BAT was tested for 20 benchmark networks, as described in Section 5.2.

**5.2 Example**

To explain the procedure of the proposed BAT for finding all the connected vectors and calculating the reliability in terms of these connected vectors, the step-by-step procedure is demonstrated using the binary-state network shown in Fig. 1.

**Solution:**

**STEP 0.** Remove $e_{2,1}$, $e_{3,1}$, $e_{2,4}$, and $e_{3,4}$ from Fig. 1; i.e., Fig. 2 is the new graph. Let $m = 5$ (in Fig. 2) and $n_c = n_p = 2$, since $\{a_1, a_2\}$ and $\{a_1, a_4\}$ are one of the minimum (1, $n$)-cuts and one of the shortest directed (1, $n$)-paths in Fig. 2 (as discussed in Section 4.3.2), respectively.

**STEP 1.** Let $R = k = 0$ and $X = (0, 0, 0, 1, 1)$; because $n_p = 2$, SUM = $n_p = 2$. Go to STEP 6.

**STEP 6.** Because SUM = $2 < m - n_c = 3$, go to STEP 7.

**STEP 7.** Because nodes 1 and $n$ are disconnected in $G(X)$ after the PLSA is used, go to STEP 2.

**STEP 2.** Let $i = 5$.

**STEP 3.** Because $X(a_5) = 1$, go to STEP 4.

**STEP 4.** Let $X(a_5) = 0$ and SUM = SUM − 1 = 1. Because $i = 5 > 1$, let $i = i - 1 = 4$ and go STEP 3.

**STEP 3.** Because $X(a_4) = 1$, go to STEP 4.

**STEP 4.** Let $X(a_4) = 0$ and SUM = SUM − 1 = 0. Because $i = 4 > 1$, let $i = i - 1 = 3$ and go STEP 3.

**STEP 3.** Because $X(a_3) = 0$, let $X(a_3) = 1$ and SUM = SUM + 1 = 1, and go STEP 5.

⋮

**STEP 2.** Let $i = 5$. Note the current $X = (1, 1, 1, 0, 0)$.

**STEP 3.** Because $X(a_5) = 0$, let $X(a_5) = 1$ and SUM = SUM + 1 = 4, and go STEP 5.

**STEP 5.** Because SUM = $4 > n_p = 2$, go to STEP 6.

**STEP 6.** Because SUM = $4 > m - n_c = 3$, go to STEP 8.

**STEP 8.** Let $k = k + 1 = 13$, $X_{13} = X = (1, 1, 1, 0, 1)$, and $R = R + p_1p_2p_3q_4p_5$, and go STEP 9.

**STEP 9.** Because SUM = 4 < 5, go to STEP 2.

⋮

The final results of the proposed BAT for finding all the connected state vectors are presented in Table 4; $R = q_1p_2q_3q_4p_5 + q_1p_2q_3p_4p_5 + q_1p_2p_3q_4p_5 + q_1p_2p_3p_4p_5 + p_1q_2q_3p_4q_5 + p_1q_2q_3p_4p_5 + p_1q_2p_3q_4p_5 + p_1q_2p_3p_4q_5 + p_1q_2p_3p_4p_5 + p_1p_2q_3q_4p_5 + p_1p_2q_3p_4q_5 + p_1p_2q_3p_4p_5 + p_1p_2p_3q_4p_5 + p_1p_2p_3p_4q_5 + p_1p_2p_3p_4p_5$.

**Table 4.** Final results of the proposed BAT for the illustrated example.

| Iterative | $k$ | $X$ | SUM | Test Method | Connected? | $R(X_k)$ |
|---|---|---|---|---|---|---|
| 1 | | (0, 0, 0, 1, 1) | 2 | PLSA | N | |
| 2 | | (0, 0, 1, 0, 0) | 1 | SUM$<n_p$ | N | |
| 3 | | (0, 0, 1, 0, 1) | 2 | PLSA | N | |
| 4 | | (0, 0, 1, 1, 0) | 2 | PLSA | N | |
| 5 | | (0, 0, 1, 1, 1) | 3 | PLSA | N | |
| 6 | | (0, 1, 0, 0, 0) | 1 | SUM$<n_p$ | N | |
| 7 | 1 | (0, 1, 0, 0, 1) | 2 | PLSA | Y | $q_1p_2q_3q_4p_5$ |
| 8 | | (0, 1, 0, 1, 0) | 2 | PLSA | N | |
| 9 | 2 | (0, 1, 0, 1, 1) | 3 | PLSA | Y | $q_1p_2q_3p_4p_5$ |
| 10 | | (0, 1, 1, 0, 0) | 2 | PLSA | N | |
| 11 | 3 | (0, 1, 1, 0, 1) | 3 | PLSA | Y | $q_1p_2p_3q_4p_5$ |
| 12 | | (0, 1, 1, 1, 0) | 3 | PLSA | N | |
| 13 | 4 | (0, 1, 1, 1, 1) | 4 | SUM$>m-n_c$ | Y | $q_1p_2p_3p_4p_5$ |
| 14 | | (1, 0, 0, 0, 0) | 1 | SUM$<n_p$ | N | |
| 15 | | (1, 0, 0, 0, 1) | 2 | PLSA | N | |
| 16 | 5 | (1, 0, 0, 1, 0) | 2 | PLSA | Y | $p_1q_2q_3p_4q_5$ |
| 17 | 6 | (1, 0, 0, 1, 1) | 3 | PLSA | Y | $p_1q_2q_3p_4p_5$ |
| 18 | | (1, 0, 1, 0, 0) | 2 | PLSA | N | |
| 19 | 7 | (1, 0, 1, 0, 1) | 3 | PLSA | Y | $p_1q_2p_3q_4p_5$ |
| 20 | 8 | (1, 0, 1, 1, 0) | 3 | PLSA | Y | $p_1q_2p_3p_4q_5$ |
| 21 | 9 | (1, 0, 1, 1, 1) | 4 | SUM$>m-n_c$ | Y | $p_1q_2p_3p_4p_5$ |
| 22 | | (1, 1, 0, 0, 0) | 2 | PLSA | N | |
| 23 | 10 | (1, 1, 0, 0, 1) | 3 | PLSA | Y | $p_1p_2q_3q_4p_5$ |
| 24 | 11 | (1, 1, 0, 1, 0) | 3 | PLSA | Y | $p_1p_2q_3p_4q_5$ |
| 25 | 12 | (1, 1, 0, 1, 1) | 4 | SUM$>m-n_c$ | Y | $p_1p_2q_3p_4p_5$ |
| 26 | | (1, 1, 1, 0, 0) | 3 | PLSA | N | |
| 27 | 13 | (1, 1, 1, 0, 1) | 4 | SUM$>m-n_c$ | Y | $p_1p_2p_3q_4p_5$ |
| 28 | 14 | (1, 1, 1, 1, 0) | 4 | SUM$>m-n_c$ | Y | $p_1p_2p_3p_4q_5$ |
| 29 | 15 | (1, 1, 1, 1, 1) | 5 | SUM$>m-n_c$ | Y | $p_1p_2p_3p_4p_5$ |

□

### 5.3 Computation Experiments

We now demonstrate the superior performance of the BAT via computation experiments involving 20 benchmark problems. The quick inclusion–exclusion method (QIE) presented in [10] is based on the IET and is one of the most efficient indirect methods [1, 4, 9, 22, 27, 33, 39] for evaluating the network reliability [10, 30]. Hence, the QIE was compared with the proposed BAT for calculating the reliability of binary-state networks.

To obtain a fair comparison, similar to the QIE [10], the proposed BAT was coded in DEV C++, run on an Intel Core i7-6650U CPU @ 2.20GHz 2.21GHz notebook with 16 GB of memory (64-bit Windows 10), and tested on 20 benchmark binary-state networks [3, 6, 9, 19], as shown in Fig. 5. These 20 networks have been widely used in previous studies to validate the performance and effectiveness of new algorithms. Note that the original Fig. 5(m) and Fig. 5(p) are drawn in directed graphs. However, in fact, their reliabilities were calculated according to all the undirected arcs in [10]. Hence, these two graphs are corrected to undirected graphs in Fig. 5.

Likewise, the reliability of each node is 0.9, and the time limit is 10 h per benchmark network; i.e., the program is forced to terminate if its runtime exceeds 10 h, as in [10]. The results of the QIE and the proposed BAT are presented in Table 5, where all the QIE results were obtained from [10]. In the first row of Table 5, the notations $n$, $m^*$, $m$, and $R$ are defined in Section 2.2; $|P|$ represents the number of MPs; $N_{QIE}$ represents the number of terms in Eq. (1) for calculating the reliability in the QIE; $N_{BAT}$ represents the total number of connected vectors obtained from the BAT; and $T_{QIE}$ and $T_{BAT}$ represent the runtimes of the QIE and BAT, respectively.

In Table 5, the bold numbers represent the best results between the proposed BAT and the QIE. As discussed in [10] (and shown in Table 5), the runtime of the QIE was zero if the number of MPs was <10 (e.g., Figs. 5(a)–4(e)), and the QIE failed to solve the benchmark networks with ≤44 MPs within 10 h (i.e., Figs. 5(m)–5(t)).

The QIE was more efficient than the BAT for $|P| \leq 20$, and $N_{QIE}$ was $<N_{BAT}$ for $|P| < 24$, as shown in Table 5; e.g., $N_{QIE} = 1048575 < N_{BAT} = 4877312$ in Fig. 5(j). However, both situations started reversely: the BAT was better than the QIE for all $|P| > 20$, and the number of items used in the BAT for calculating the reliability was smaller than that used in the QIE for $|P| \geq 24$. The difference became remarkable after $|P| \geq 29$: the runtime was only 0.10500000 for the BAT but was 21.75000000 for the QIE for Fig. 5(l). Moreover, the proposed BAT was able to solve the problems shown in Figs. 5(o) and 5(p), in contrast to the QIE. Additionally, the BAT failed for Figs. 5(m), (n), and (q)–(t); this obstacle is very difficult to overcome, because calculating the network reliability is an NP-hard problem.

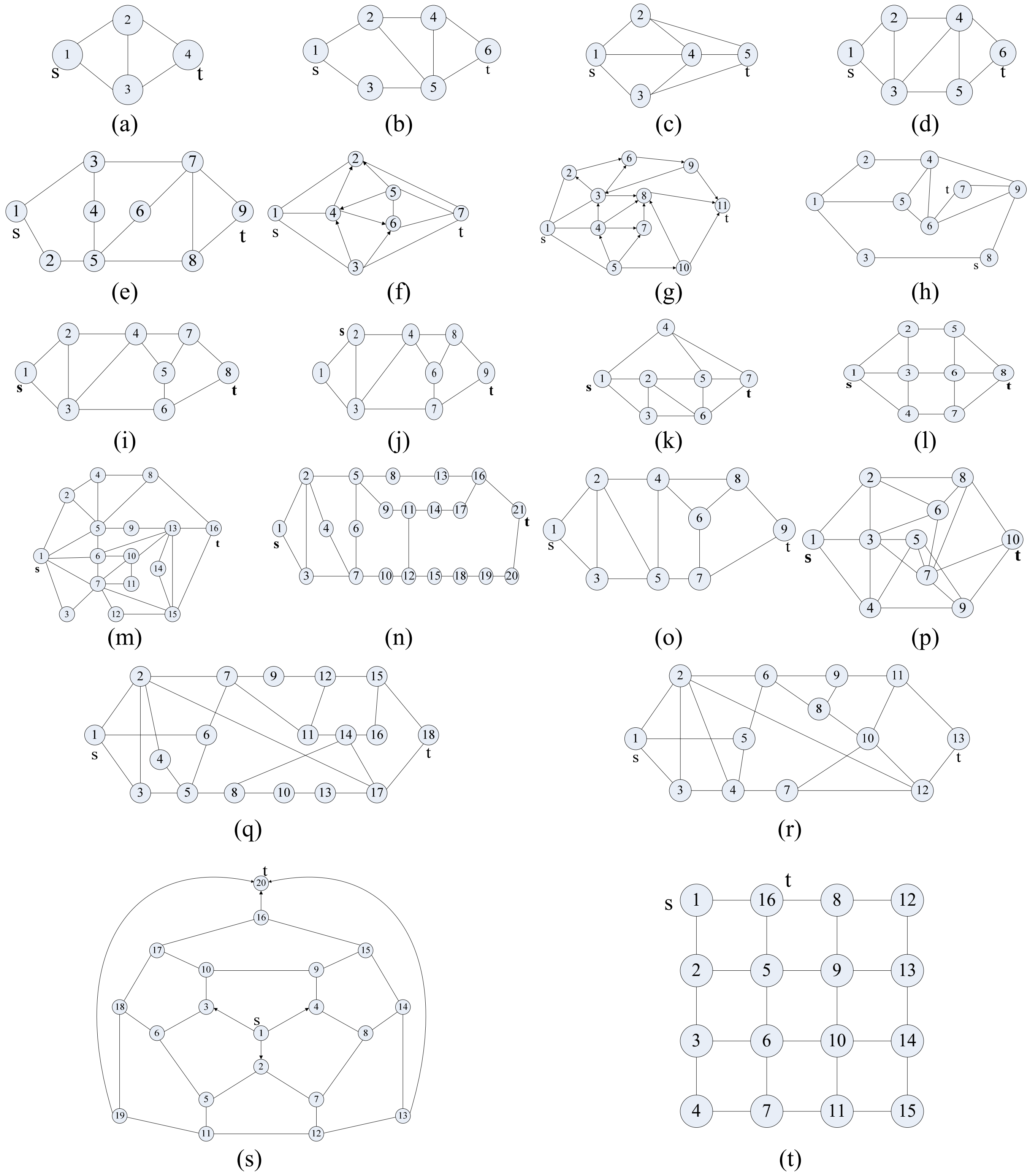

**Figure 5.** 20 benchmark binary-state networks used in the test

**Table 5.** Comparison of the BAT and QIE [10].

| Fig. | *n* | *m** | *m* | $n_p$ | $n_c$ | \|*P*\| | $N_{QIE}$ | $N_{BAT}$ | $T_{QIE}$ | $T_{BAT}$ | *R* |
|---|---|---|---|---|---|---|---|---|---|---|---|
| a | 4 | 10 | 10 | 2 | 2 | 4 | **15** | 32 | 0.00000000 | 0.00000000 | 0.9784800000 |
| b | 6 | 16 | 16 | 3 | 2 | 7 | **127** | 1456 | 0.00000000 | 0.00000000 | 0.9684254700 |
| c | 5 | 16 | 16 | 2 | 3 | 9 | **511** | 688 | 0.00000000 | 0.00000000 | 0.9976316400 |
| d | 6 | 18 | 18 | 3 | 2 | 13 | 8191 | **6560** | **0.00000000** | 0.00200000 | 0.9771844050 |
| e | 9 | 24 | 24 | 3 | 2 | 13 | **8191** | 277760 | **0.00000000** | 0.08400000 | 0.9648551232 |
| f | 7 | 22 | 22 | 2 | 3 | 14 | **16383** | 20096 | 0.00100000 | **0.00000000** | 0.9966644040 |
| g | 11 | 25 | 25 | 3 | 3 | 18 | **262143** | 991522 | **0.01600000** | 0.23199999 | 0.9940757879 |
| h | 9 | 26 | 26 | 2 | 2 | 18 | **262143** | 1574912 | **0.01100000** | 0.28600000 | 0.9691117946 |
| i | 8 | 24 | 24 | 3 | 2 | 24 | 16777215 | **335104** | 0.62099999 | **0.08200000** | 0.9751158974 |
| j | 9 | 28 | 28 | 4 | 2 | 20 | **1048575** | 4877312 | **0.03900000** | 1.35099995 | 0.9840681530 |
| k | 7 | 24 | 24 | 2 | 3 | 25 | 33554431 | **158208** | 1.24000001 | **0.02400000** | 0.9974936737 |
| l | 8 | 26 | 26 | 3 | 3 | 29 | 536870911 | **524288** | 21.75000000 | **0.10500000** | 0.9962174933 |
| m | 10 | 18 | 18 | 3 | 3 | 257 | 2.31584E+77 | | | | |
| n | 21 | 52 | 52 | 6 | 2 | 44 | 1.75922E+13 | | | | |
| o | 9 | 28 | 28 | 4 | 2 | 44 | 1.75922E+13 | **4708352** | | 1.11399996 | 0.9741454748 |
| p | 10 | 42 | 42 | 3 | 3 | 331 | 4.3745E+99 | **44719538176** | | 6950.24023438 | 0.9979623223 |
| q | 18 | 54 | 54 | 3 | 2 | 269 | 9.48569E+80 | | | | |
| r | 13 | 44 | 44 | 3 | 2 | | | | | | 3,2 |
| s | 20 | 60 | 60 | 5 | 3 | | | | | | 5,3 |
| t | 16 | 48 | 48 | 1 | 2 | | | | | | |

According to the foregoing experimental results, the BAT is more attractive than the QIE for middle-size networks. These results confirm the conclusions based on the time complexity presented in Section 5.1.

## 6. CONCLUSIONS

This paper presents a novel direct method called the BAT that employs a binary-addition tree to find all the state vectors, the PLSA to find the connected state vectors, a method to calculate the reliability according to the connected vectors, and reduction methods to reduce the computational burden.

MC and MP algorithms both play significant roles in assessing the reliability of binary-state networks. However, this study revealed that these algorithms are less efficient than the proposed BAT for evaluating the reliability of a binary-state network. Regarding the time complexity, the proposed BAT was significantly better than the best-known MP/MC algorithms [27, 33], which need to have all the MPs/MCs first and implement the IET or SDP to calculate the reliability in terms of the MPs/MCs.

In a computation experiment involving 20 benchmark problems, the performance of the proposed BAT was superior to that of the best-known indirect algorithm proposed in [10]. Moreover, the proposed BAT based on binary addition is easier to understand and implement than the DFS, universal generating function method [15, 16, 31], heuristic [29], etc. used in the traditional indirect methods [4, 9, 22, 27, 33, 39], for example, MP/MC algorithms [4, 9, 22, 27, 33, 39] or the direct method, e.g., BBD [17].

Thus, from a general, practical, and theoretical viewpoint, the proposed BAT is more attractive than the existing algorithms for calculating the reliability of binary-state networks.

## ACKNOWLEDGMENT

This research was supported in part by the Ministry of Science and Technology, R.O.C. under grant MOST 107-2221-E-007-072-MY3.

## REFERENCES

[1] Ahmad, S. H. (1988). "Simple enumeration of minimal cutsets of acyclic directed graph." IEEE transactions on reliability **37**(5): 484-487.

[2] Aven, T. (1987). "Availability evaluation of oil/gas production and transportation systems." Reliability engineering **18**(1): 35-44.

[3] Aven, T. (1988). "Some considerations on reliability theory and its applications." Reliability Engineering & System Safety **21**(3): 215-223.

[4] Bai, G., M. J. Zuo and Z. Tian (2015). "Search for all *d*-MPs for all *d* levels in multistate two-terminal networks." Reliability Engineering & System Safety **142**: 300-309.

[5] Bhavathrathan, B. and G. R. Patil (2013). "Analysis of worst case stochastic link capacity degradation to aid assessment of transportation network reliability." Procedia-Social and Behavioral Sciences **104**: 507-515.

[6] Bryant, R. E. (1986). "Graph-based algorithms for boolean function manipulation." Computers, IEEE Transactions on **100**(8): 677-691.

[7] Coit, D. W. and E. Zio (2018). "The evolution of system reliability optimization." Reliability Engineering & System Safety: 106259.

[8] Colbourn, C. J. (1987). The combinatorics of network reliability, Oxford University Press, Inc.

[9] Forghani-Elahabad, M. and N. Kagan (2019). "Assessing reliability of multistate flow networks under cost constraint in terms of minimal cuts." International Journal of Reliability, Quality and Safety Engineering **26**(5): 1950025.

[10] Hao, Z., W.-C. Yeh, J. Wang, G.-G. Wang and B. Sun (2019). "A quick inclusion-exclusion technique." Information Sciences **486**: 20-30.

[11] Hwang, F. (1991). "Reliability of computer and communication networks." DIMACS Amer Math Soc.

[12] Kakadia, D. and J. E. Ramirez-Marquez (2020). "Quantitative approaches for optimization of user experience based on network resilience for wireless service provider networks." Reliability Engineering & System Safety **193**: 106606.

[13] Laitrakun, S. and E. J. Coyle (2014). "Reliability-based splitting algorithms for time-constrained distributed detection in random-access WSNs." IEEE Transactions on Signal Processing **62**(21): 5536-5551.

[14] Lee, C.-Y. (1959). "Representation of switching circuits by binary-decision programs." The Bell System Technical Journal **38**(4): 985-999.

[15] Levitin, G. (2005). The universal generating function in reliability analysis and optimization, Springer.

[16] Levitin, G., L. Xing and Y. Dai (2017). "Optimal Spot-Checking for Collusion Tolerance in Computer Grids." IEEE Transactions on Dependable and Secure Computing **16**(2): 301-312.

[17] Lin, C., L. Cui, D. W. Coit and M. Lv (2017). "Performance analysis for a wireless sensor network of star topology with random nodes deployment." Wireless Personal Communications **97**(3): 3993-4013.

[18] Niu, Y., Z. Gao and H. Sun (2017). "An improved algorithm for solving all $d$-MPs in multi-state networks." Journal of Systems Science and Systems Engineering **26**(6): 711-731.

[19] Niu, Y.-F. and F.-M. Shao (2011). "A practical bounding algorithm for computing two-terminal reliability based on decomposition technique." Computers & Mathematics with Applications **61**(8): 2241-2246.

[20] Ramirez-Marquez, J. E. (2015). "Assessment of the transition-rates importance of Markovian systems at steady state using the unscented transformation." Reliability Engineering & System Safety **142**: 212-220.

[21] Sanseverino, C. M. R. and J. E. Ramirez-Marquez (2014). "Uncertainty propagation and sensitivity analysis in system reliability assessment via unscented transformation." Reliability Engineering & System Safety **132**: 176-185.

[22] Shier, D. R. (1991). Network reliability and algebraic structures, Clarendon Press.

[23] Song, Y. and B. Wang (2012). "Survey on reliability of power electronic systems." IEEE Transactions on Power Electronics **28**(1): 591-604.

[24] Wang, J., W.-C. Yeh, N. N. Xiong, J. Wang, X. He and C.-L. Huang (2019). "Building an improved Internet of things smart sensor network based on a three-phase methodology." IEEE Access **7**: 141728-141737.

[25] Wang, Y., L. Xing, H. Wang and D. W. Coit (2017). "System reliability modeling considering correlated probabilistic competing failures." IEEE Transactions on Reliability **67**(2): 416-431.

[26] Yeh, W.-C. (1998). "A revised layered-network algorithm to search for all d-minpaths of a limited-flow acyclic network." IEEE Transactions on Reliability **47**(4): 436-442.

[27] Yeh, W.-C. (2006). "A simple algorithm to search for all MCs in networks." European Journal of Operational Research **174**(3): 1694-1705.

[28] Yeh, W.-C. (2007). "An improved sum-of-disjoint-products technique for the symbolic network reliability analysis with known minimal paths." Reliability Engineering & System Safety **92**(2): 260-268.

[29] Yeh, W.-C. (2007). "A simple heuristic algorithm for generating all minimal paths." IEEE Transactions on Reliability **56**(3): 488-494.

[30] Yeh, W.-C. (2008). "A greedy branch-and-bound inclusion-exclusion algorithm for calculating the exact multi-state network reliability." IEEE Transactions on Reliability **57**(1): 88-93.

[31] Yeh, W.-C. (2009). "A simple universal generating function method to search for all minimal paths in networks." IEEE Transactions on Systems, Man, and Cybernetics-Part A: Systems and Humans **39**(6): 1247-1254.

[32] Yeh, W.-C. (2015). "An improved sum-of-disjoint-products technique for symbolic multi-state flow network reliability." IEEE Transactions on Reliability **64**(4): 1185-1193.

[33] Yeh, W.-C. (2016). "New method in searching for all minimal paths for the directed acyclic network reliability problem." IEEE Transactions on Reliability **65**(3): 1263-1270.

[34] Yeh, W.-C. (2016). "A squeezed artificial neural network for the symbolic network reliability functions of binary-state networks." IEEE transactions on neural networks and learning systems **28**(11): 2822-2825.

[35] Yeh, W.-C. (2019). "A Novel Generalized Artificial Neural Network for Mining Two-Class Datasets." arXiv preprint arXiv:1910.10461.

[36] Yeh, W.-C. and J.-S. Lin (2018). "New parallel swarm algorithm for smart sensor systems redundancy allocation problems in the Internet of Things." The Journal of Supercomputing **74**(9): 4358-4384.

[37] Yeh, W.-C. and S.-C. Wei (2012). "Economic-based resource allocation for reliable Grid-computing service based on Grid Bank." Future Generation Computer Systems **28**(7): 989-1002.

[38] Zhang, L., H. Ma, D. Shi, P. Wang, G. Cai and X. Liu (2017). "Reliability oriented modeling and analysis of vehicular power line communication for vehicle to grid (V2G) information exchange system." IEEE Access **5**: 12449-12457.

[39] Zuo, M. J., Z. Tian and H.-Z. Huang (2007). "An efficient method for reliability evaluation of multistate networks given all minimal path vectors." IIE transactions **39**(8): 811-817.